\documentclass[%
 reprint,
 amsmath,amssymb,
 aps,
]{revtex4-2}

\usepackage{graphicx}
\usepackage{dcolumn}
\usepackage{bm}
\usepackage{physics}
\usepackage{qcircuit}
\usepackage{amsmath}
\usepackage{caption}
\usepackage{subcaption}
\usepackage{CJK}
\usepackage{mathrsfs}
\usepackage{amsfonts}
\usepackage{mathtools}
\usepackage{braket}

\usepackage{xcolor}

\def\be{\begin{equation}}
\def\ee{\end{equation}}
\def\ba{\begin{eqnarray}}
\def\ea{\end{eqnarray}}

\begin{document}

\title{Lorentz Quantum Computer}

\begin{CJK}{UTF8}{gbsn}
\author{Wenhao He(何文昊)}
\affiliation{International Center for Quantum Materials, School of Physics, Peking University,  Beijing 100871, China}
\author{Zhenduo Wang(王朕铎)}
\affiliation{International Center for Quantum Materials, School of Physics, Peking University,  Beijing 100871, China}
\author{Biao Wu(吴飙)}
\email{wubiao@pku.edu.cn}
\affiliation{International Center for Quantum Materials, School of Physics, Peking University,  Beijing 100871, China}
\affiliation{Wilczek Quantum Center, School of Physics and Astronomy, Shanghai Jiao Tong University, Shanghai 200240, China}
\affiliation{Collaborative Innovation Center of Quantum Matter, Beijing 100871,  China}


\begin{abstract}
A theoretical model of computation is proposed based on Lorentz quantum mechanics.  Besides the standard qubits, 
this model has an additional bit, which we call hyperbolic bit (or hybit in short).  A set of basic logical gates 
are constructed and their universality is proved. As an application, a search algorithm is designed  for
this computer model and is found to be exponentially faster than the Grover's search algorithm. 
\end{abstract}

\maketitle
\end{CJK}

\section{Introduction}
The theoretical models of computation had long been regarded mistakenly as a pure mathematical structure. 
This view was completely changed with the rise of quantum computer.  This is well summarized 
by Deutsch~\cite{Deustch}, ``computers are physical objects,$\cdots$, what computers can or cannot 
do is determined by laws of physics alone''.  In other words, different physical theories lead to 
different computational models with distinct computing powers. 

Currently, there are only two 
well established frameworks of mechanics, classical mechanics (including Maxwell equations and general relativity) and 
quantum mechanics (including quantum field theories). Consequently,  there are two types of computers, classical computer
and quantum computer. It is naturally to conjecture new kinds of mechanics, and use it as a base to establish new 
models of computers.

We are going to discuss a computational model based on Lorentz quantum mechanics, 
where the dynamical evolution is  complex Lorentz transformation. 
It  was proposed in  Ref.\cite{zhang2018lorentz} as a generalization of the Bogoliubov-de Gennes equation; 
similar mechanics was  studied a long time ago by Pauli~\cite{Pauli1943}. 
The key feature in Lorentz mechanics, which has an indefinite metric,  is that only 
the states with the positive norms are physically observable. 

We introduce a  bit called 
hyperbolic bit (or hybit in short). The Lorentz computer so established consists of  both qubits and hybits, which are 
manipulated by a set of basic logical gates. The universality of these gates are rigorously proved.  By construction quantum computer
is a special case of Lorentz computer, we thus expect the Lorentz computer to be more powerful. 
This is indeed the case as we find a Lorentz search algorithm that is more powerful than the Grover's search 
algorithm~\cite{Grover}.  We will discuss the physical implementation of our computer model
as a  single Lorentz system was recently  simulated with photons  with postselection~\cite{Li2019}.

\section{Lorentz Quantum Mechanics}
Lorentz quantum mechanics was discussed in ~\cite{zhang2018lorentz} as a generalization of the
Bogoliubov-de Gennes  equation~\cite{zhang2018lorentz,wunjp,Artem}.  However, 
this kind of generalized quantum mechanics with indefinite metric was  studied a long time ago 
by  Pauli~\cite{Pauli1943}. The related mathematical structure has been studied systematically~\cite{Bognar}. 
In the following, we briefly review the framework of Lorentz quantum mechanics, and then introduce 
the composite systems consisting of  quantum systems and Lorentz systems,  which are the key to our computer model. 

\subsection{General theory of Lorentz systems}
Quantum states are vectors $\ket{\psi}$ in a Hilbert space where the inner product $\braket{\psi|\psi}$ is always non-negative. 
The states $|\psi)$ of Lorentz quantum mechanics  are vectors of a linear space with inner product defined as
$(\psi|\eta|\psi)$, which can be negative. The indefinite metric $\eta$ is a Hermitian matrix. When $\eta$ is an identity matrix, 
we recover the usual Hilbert space. In our following notation, when $|~~)$ is used, $\eta$ is  not  an identity matrix; 
when $|~~\rangle$ is used, $\eta$ is  an identity matrix. 
The general form of Lorentz quantum mechanics is given by 
\be
\mathrm{i} \frac{\mathrm{d}}{\mathrm{d} t}|\psi)=\eta \hat{H}|\psi)\,,
\label{general}
\ee
where $\hat{H}$ is a Hermitian Hamiltonian. 

We focus our attention to the special case $\eta=\eta_{m,n}$, where~\cite{zhang2018lorentz}
\begin{eqnarray}
\eta_{m, n}=\operatorname{diag}\{\underbrace{1,1, \ldots 1}_{m}, \underbrace{-1,-1, \ldots-1}_{n}\}\,.
\end{eqnarray}
When $m=n$, Eq.(\ref{general}) becomes the well-known Bogoliubov-de Gennes  equation~\cite{zhang2018lorentz,wunjp}. 
The evolution operator $\hat{\mathcal{U}}$ can be written as
\begin{eqnarray}
\hat{\mathcal{U}}(t, 0)=\mathrm{e}^{-\mathrm{i} \eta_{m,n} \hat{H} t}\,.
\label{eq:evolution}
\end{eqnarray}
It can be verified that $\hat{\mathcal{U}}^{\dagger} \eta_{m,n} \hat{\mathcal{U}}=\eta_{m,n}$. 
This means that $\hat{\mathcal{U}}$ is a member of the generalized Lorentz group $U(m,n)$. 
And we call $\hat{\mathcal{U}}$ complex Lorentz transformation,  under which
the norm of $|\psi)$ is conserved, 
\begin{eqnarray}
	\frac{\mathrm{d}}{\mathrm{d} t}(\psi| \eta_{m,n} |\psi) = 0\,.
\label{eq:prob_cons}
\end{eqnarray}
The complex Lorentz transformation $\hat{\mathcal{U}}$ is called isometric operator in Ref.~\cite{Bognar}.
When $n\neq 0$, the norm can be positive, negative, and zero.  The positive one will be normalized to one
and the negative one be normalized to minus one.   

\subsection{Single Lorentz systems}
There are two kinds of  quantum systems, single systems and  composite systems. For example, 
although both of their Hilbert spaces are of dimension four, spin-3/2 is a single system and the system with two spin-1/2's 
is a composite system. Similarly, there are also two kinds of Lorentz systems, single systems and composite systems. 
In single Lorentz systems,  only the states $|\phi)$ with positive norm $(\phi|\eta_{m,n}|\phi) > 0$ are regarded 
as physical and observable.

A quantum state $\ket{\psi}$ is not directly observable in general. In a measurement regarding operator $A$ whose
eigenstates are $\ket{\psi_j}$'s, one
observes $\ket{\psi_j}$ with probability $|\braket{\psi_j|\psi}|^2$.  
Similarly, a Lorentz state $|\phi)$ is not directly observable in general. Consider a Lorentz operator ${\bm B}$, which is not 
necessarily Hermitian, with  eigenstates $\{|\phi_j), j=1,2,\cdots,m\}$ and 
$\{|\varphi_j), j=1,2,\cdots,n\}$. We assume that $(\phi_j|\eta_{m,n}|\phi_j)=1$ and $(\varphi_j|\eta_{m,n}|\varphi_j)=-1$.
With the expansion 
\be
|\phi)=\sum_{j=1}^m a_j|\phi_j)+\sum_{j=1}^n b_j|\varphi_j)\,,
\ee
one observes $|\phi_j)$ with probability $|a_j|^2/(\sum_{j=1}^m |a_j|^2)$. The $|\varphi_j)$'s are not observable. 

It is well known in the field of superfluidity, the Bogoliubov-de Gennes  equation has two sets of eigenmodes, 
one half of them have positive norm and the other half have negative norm. The positive ones are quasi-particles
of a superfluid, such as phonons, and can be observed in experiment while the negative half are unphysical 
and have never been observed~\cite{wunjp,BEC}.  However, in the dynamics governed by the Bogoliubov-de Gennes  equation, 
these two modes are mixed together and must be taken into account simultaneously to describe some phenomena, such as 
the transverse force acting on a vortex~\cite{Artem}. This is the physical basis for 
the features of a single  Lorentz system discussed above.

\subsection{Composite systems}
We consider two basic composite systems, one consisting of a  quantum system $S_q$ and a single Lorentz system $S_l$ and 
the other consisting of two single Lorentz systems $S_{l1}$ and $S_{l2}$.  Other composite systems can be readily derived from them.  

For the first kind of composite system,
if the metric of $S_q$ is $\eta_{m_1,0}$ and the metric of $S_l$  is $\eta_{m_2,n}$, then the metric for
the composite system $S_a$ is  $\eta_{m_1,0}\otimes \eta_{m_2,n}$. Namely, the composite system is also a
Lorentz system with indefinite metric. If  the Hilbert space of $S_q$ is spanned by $\{\ket{\psi_{q}^{(i)}}, i=1,2,\cdots,m_1\}$ 
and the inner product space of $S_a$ spanned by $\{|\phi_{l}^{(j)}),j=1,2,\cdots,m_2\}$ and $\{|\varphi_{l}^{(j)}),j=1,2,\cdots,n\}$, 
then the composite system $S_a$ is spanned by
$\ket{\psi_{q}^{(i)}}\otimes |\phi_{l}^{(j)})$ and $\ket{\psi_{q}^{(i)}}\otimes |\varphi_{l}^{(j)})$. 
For a state $|\Phi)$ of the composite system $S_a$, it can be expanded as 
\be
|\Phi)=\sum_{i=1}^{m_1}\Big\{\sum_{j=1}^{m_2} a_{ij}\ket{\psi_{q}^{(i)}}\otimes |\phi_{l}^{(j)})+
\sum_{j=1}^{n} b_{ij}\ket{\psi_{q}^{(i)}}\otimes |\varphi_{l}^{(j)})\Big\}\,.
\ee
The probability of observing $\ket{\psi_{q}^{(i)}}\otimes |\phi_{l}^{(j)})$ is 
\be
P_{ij}=\frac{|a_{ij}|^2}{\sum_{i=1}^{m_1}\sum_{j=1}^{m_2}|a_{ij}|^2}\,.
\ee
And $\ket{\psi_{q}^{(i)}}\otimes |\varphi_{l}^{(j)})$ can not be observed. 

For the second kind of composite system,
if the metric of $S_{l1}$ is $\eta_{m_1,n_1}$ and the metric of $S_{l2}$  is $\eta_{m_2,n_2}$, then the metric for
the composite system $S_b$ is  $\eta_{m_1,n_1}\otimes \eta_{m_2,n_2}$. 
If  the Hilbert space of $S_{l1}$ is spanned by 
$\{|\phi_{l1}^{(j)}),j=1,2,\cdots,m_1\}$ and $\{|\varphi_{l1}^{(j)}),j=1,2,\cdots,n_1\}$, 
and the inner product space of $S_{l2}$ spanned by $\{|\phi_{l2}^{(j)}),j=1,2,\cdots,m_2\}$ and $\{|\varphi_{l2}^{(j)}),j=1,2,\cdots,n_2\}$, 
then the composite system $S_b$ is spanned by
$|\phi_{l1}^{(j_1)})\otimes |\phi_{l2}^{(j_2)})$, $|\phi_{l}^{(j_1)})\otimes |\varphi_{l2}^{(j_2)})$, 
$|\varphi_{l1}^{(j_1)})\otimes |\phi_{l2}^{(j_2)})$, and $|\varphi_{l1}^{(j_1)})\otimes |\varphi_{l2}^{(j_2)})$. 
For a state $|\Phi)$ of the composite system $S_b$, it can be expanded as 
\ba
|\Phi) &=&\sum_{j_1=1}^{m_1}\sum_{j_2=1}^{m_2} a_{j_1j_2}|\phi_{l1}^{(j_1)})\otimes |\phi_{l2}^{(j_2)})+\nonumber\\
&&\sum_{j_1=1}^{m_1}\sum_{j_2=1}^{n_2} b_{j_1j_2}|\phi_{l1}^{(j_1)})\otimes |\varphi_{l2}^{(j_2)})+\nonumber\\
&&\sum_{j_1=1}^{n_1}\sum_{j_2=1}^{m_2} c_{j_1j_2}|\varphi_{l1}^{(j_1)})\otimes |\phi_{l2}^{(j_2)})+\nonumber\\
&&\sum_{j_1=1}^{n_1}\sum_{j_2=1}^{n_2} d_{j_1j_2}|\varphi_{l1}^{(j_1)})\otimes |\varphi_{l2}^{(j_2)})\,.
\ea
The probability of observing $|\phi_{l1}^{(j_1)})\otimes |\phi_{l2}^{(j_2)})$ is 
\be
P_{j_1j_2}=\frac{|a_{j_1j_2}|^2}{\sum_{j_1=1}^{m_1}\sum_{j_2=1}^{m_2}|a_{j_1j_2}|^2}\,.
\ee
And other states $|\phi_{l1}^{(j_1)})\otimes |\varphi_{l2}^{(j_2)})$, $|\varphi_{l1}^{(j_1)})\otimes |\phi_{l2}^{(j_2)})$, and 
$|\varphi_{l1}^{(j_1)})\otimes |\varphi_{l2}^{(j_2)})$ are not observable. 

\section{Model of Lorentz Computing}
There are three essential parts in a computational model:
\begin{enumerate}
\item What represents information? (encoding)
\item How  is the information processed? (computing) 
\item How to extract the information? (decoding) 
\end{enumerate}

In a classical computer, the information is stored in bits. The information is processed with classical logical gates. 
For reversible classical computer, either the Fredkin gate or the Toffoli gate can serve as 
the universal gate~\cite{nielson2000quantum}. 
At the end of computation, the output is  recorded in a string of bits with each bit in a definite state, 0 or 1. 

In a quantum computer, the information is stored in qubits and 
the information is processed with quantum logical gates. 
There are three universal gates, Hadamard gate $\hat{H}$, $\pi/8$ gate $\hat{T}$, and CNOT gate. 
These gates are unitary transformations. 
At the end of computation, the qubits are usually in a superposition state where each qubit is  
not in a definite state.  A measurement is then performed so that each qubit falls into a definite state,
$\ket{0}$ or $\ket{1}$ ;
the result is then extracted~\cite{nielson2000quantum}. 

People now have realized that the classical computer embodies classical mechanics and the quantum computer
is derived from quantum mechanics. It is thus natural to construct a computer model based Lorentz quantum mechanics 
discussed above. 

In a Lorentz quantum computer, the information is stored in both qubits and hybits. This means that the Lorentz computer
is a composite system consisting of both quantum systems and Lorentz systems. The information
is then processed with a set of universal Lorentz quantum gates, which will be presented in the next section. 
These universal gates are complex Lorentz transformations and the usual quantum universal gates 
are  a subset.
At the end of computation, 
as in quantum computers, the qubits and hybits are in general in a superposition state where each qubit
or hybit is not in a definite state. A measurement is then performed to extract information. 
The essential difference is that only states with $|0)$'s are observable.

It is clear  by construction that the usual quantum computer is a special case of Lorentz computer when 
hybits are not used. This means that the Lorentz computer is potentially more powerful than quantum computer. 
This is indeed the case.  An algorithm of Lorentz computer is designed for random search; it is exponentially faster 
than the Grover algorithm~\cite{nielson2000quantum,Grover}.

\subsection{Hybits}
There are two kinds of bits in a Lorentz computer. The first is the familiar qubits, 
whose computational basis is made of  $\ket{0}$ and $\ket{1}$.  The second is unique to 
Lorentz computer and we call it hyperbolic bit (or hybit for short) as its state vector stays on a hyperbolic surface. 
For a  hybit, its general state is represented as 
\be
|\psi) = a|0) + b|1)=
\left(\begin{array}{c}
      a\\
      b\\
   \end{array}
   \right) 
\ee 
where  $|0)$ and  $|1)$ are the computational basis satisfying
\be
   (0|\eta_{1,1}|0) = 1 \,, ~~
   (1|\eta_{1,1}|1) = -1 \,, ~~
   (1|\eta_{1,1}|0) = 0 \,.
\ee
For a Lorentz computer made of $N_q$ qubits and $N_h$ hybits, its state space is spanned by direct products of 
$
 \ket{\psi_1}\otimes\cdots\otimes\ket{\psi_{N_q}}\otimes|\psi_1)\otimes\cdots\otimes|\psi_{N_h})
$
with dimension $2^{N_q+N_h}$. Such a computer is a composite system with the following metric 
\be
 \eta = \underbrace{\eta_{2,0}\otimes\cdots\otimes\eta_{2,0}}_{N_q}\otimes
 \underbrace{\eta_{1,1}\otimes\cdots\otimes\eta_{1,1}}_{N_h}\,.
\ee
The state $|\Phi)$ of a Lorentz computer can be expanded in the computational basis 
\be
|\Phi)=\sum_{j=1}^{2^{N_q+N_h}}a_j|\psi_j)\,,
\ee
where 
\ba
|\psi_j)&=&\ket{d_1}\otimes\ket{d_2}\cdots\ket{d_{N_q}}\otimes|d_1)\otimes|d_2)\cdots\otimes|d_{N_h})\nonumber\\
&=&\ket{d_1,d_2\cdots,d_{N_q},\bar{d}_1,\bar{d}_2,\cdots,\bar{d}_{N_h}}\,.
\ea
with $d_j$ and $\bar{d}_j$ being either 0 or 1. In the above and from now on, for simplicity, we used and will use
$\ket{\bar{0}}=|0)$ and $\ket{\bar{1}}=|1)$. According to Lorentz quantum mechanics introduced 
in the last section, any component with just one $|1)$, e.g., $\ket{1,0\cdots,0,\bar{1},\bar{0},\cdots,\bar{0}}$,  
is not observable. 

The simplest Lorentz computer consists of one qubit and one hybit. Its general state is given by
\be
|\Phi)=a_1\ket{0,\bar{0}}+a_2\ket{0,\bar{1}}+a_3\ket{1,\bar{0}}+a_4\ket{1,\bar{1}}\,.
\ee
Upon measurement, the probability of observing $\ket{0,\bar{0}}$ is 
$\frac{\absolutevalue{a_1}^2}{\absolutevalue{a_1}^2+\absolutevalue{a_3}^2}$
while the probability of  $\ket{1,\bar{0}}$ is $\frac{\absolutevalue{a_3}^2}{\absolutevalue{a_1}^2+\absolutevalue{a_3}^2}$.
The other two states, $\ket{0,\bar{1}}$ and $\ket{1,\bar{1}}$,  are  not  observable.

Note that the case of 
$N_h = 0$ is completely equivalent to  quantum computer. In other words, quantum computer
is a special case of Lorentz computer just as reversible classical computer is a special case of quantum computer.

\subsection{Universal Gates}
With qubits and hybits, we are ready to design logical gates for Lorentz computer. 
Similar to quantum computer, 
there also exists a set of universal gates for
Lorentz computer. The universality, which we shall  prove in the Appendix, ensures that any operator can 
be approximated to an arbitrary precision with the universal gates. 
In other words, one can use these gates to construct a set of complex Lorentz transformations,
 which is a dense set in all complex Lorentz transformations.

In the following text, $\hat{\sigma}_x , \hat{\sigma}_y$ and $\hat{\sigma}_z$ are the standard Pauli operators 
with the following matrices
\begin{eqnarray}
	\left(\begin{array}{cc}
		0 & 1 \\
		1 & 0
	\end{array}\right) \,,
	\quad
	\left(\begin{array}{cc}
		0 & -i \\
		i & 0 
	\end{array}\right) \,,
	\quad
	\left(\begin{array}{cc}
		1 & 0 \\
		0 & -1
	\end{array}\right) \,,
\end{eqnarray}
respectively.

We find that the Lorentz universal gates can be divided into three sets, 
$\{ \hat{H}, \hat{T}\}$, 
$\{ \hat{\tau}, \hat{T}\}$, and $\{\hat{\Lambda}_{1}^{qq}\left(\hat{\sigma}_{z}\right), \hat{\Lambda}_{1}^{ql}\left(\hat{\sigma}_{z}\right), 
\hat{\Lambda}_{1}^{lq}\left(\hat{\sigma}_{z}\right) ,\hat{\Lambda}_{1}^{ll}\left(\hat{\sigma}_{z}\right)\}$.
The first set $\{ \hat{H}, \hat{T}\}$ consists of Hadamard gate $\hat{H}$ and $\pi/8$ gate $\hat{T}$
\begin{eqnarray}
	\hat{H} &=& \frac{1}{\sqrt{2}}\left(\hat{\sigma}_{x}+\hat{\sigma}_{z}\right)\,,\\
\hat{T} &=& e^{-i\frac{\pi}{8}} \left(\begin{array}{cc}
	e^{i\pi/8} & 0 \\
	0 & e^{-i\pi/8} 
	\end{array}\right) \,.
\end{eqnarray}
These two operators are single qubit universal, which means that they operate on single qubits and the combination of these two gates
can approximate any single qubit transformation to an arbitrary precision.  They are denoted by symbols in Fig.\ref{QFT_circuit}. 
\begin{figure}[h!] 
	\centerline{
  		\Qcircuit @C=1.2em @R=1.1em {
			\lstick{\ket{\psi}}  & \qw & \qw  &   \gate{H}  & \qw & \qw & \qw \\	 
	 		\lstick{\ket{\psi}}  & \qw & \qw  &   \gate{T}  & \qw & \qw & \qw 
	 	}
  	}
	\caption{Symbols for single qubit gates $\hat{H}$ and $\hat{T}$.}
	\label{QFT_circuit}
\end{figure}
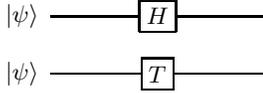

The second set operates on single hybits, 
consisting of $\pi/8$ gate $\hat{T}$ and gate $\hat{\tau}$ 
\begin{eqnarray}
	\hat{\tau} = \sqrt{2}\hat{\sigma}_{z}+i\hat{\sigma}_{x} = \left(\begin{array}{cc}
		\sqrt{2} & i \\
		i & -\sqrt{2}
		\end{array}\right) \,.
\end{eqnarray}
It can be verified that $\hat{\tau}^{\dagger}\eta_{1,1}\hat{\tau} = \eta_{1,1}$, $(\hat{T})^{\dagger}\eta_{1,1}\hat{T} = \eta_{1,1}$. These two gates are single hybit universal.
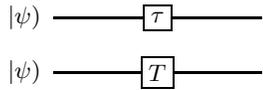
\begin{figure}[h!] 
	\centerline{
  		\Qcircuit @C=1.2em @R=1.1em {
			\lstick{|\psi)}  & \qw & \qw  &   \gate{\tau}  & \qw & \qw & \qw \\	 
	 		\lstick{|\psi)}  & \qw & \qw  &   \gate{T}  & \qw & \qw & \qw 
	 	}
  	}
	\caption{Symbols for single hybit gates $\hat{\tau}$ and $\hat{T}$.}
	\label{QFT_circuit}
\end{figure}

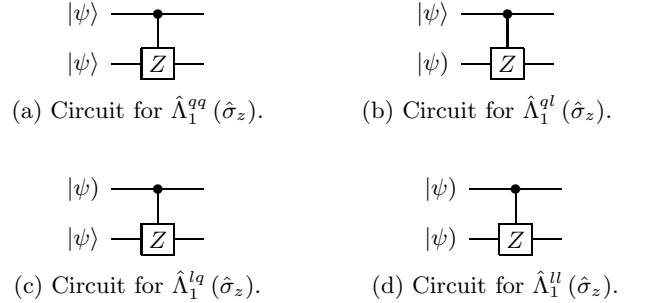
\begin{figure}[ht]
	\begin{subfigure}[b]{0.45\columnwidth}
	\centering{
	~~~~~\Qcircuit @C=1.2em @R=1.3em {
				\lstick{\ket{\psi}}   &   \ctrl{1}     &   \qw      \\
				 \lstick{\ket{\psi}}   &   \gate{Z}   &    \qw       \\			 	
			 }
	\caption{Circuit for $\hat{\Lambda}_{1}^{qq}\left(\hat{\sigma}_{z}\right)$.}}
	\end{subfigure}
	\hfill
	\begin{subfigure}[b]{0.45\columnwidth}
	~~~~~\Qcircuit @C=1.2em @R=1.3em {
				\lstick{\ket{\psi}}   &   \ctrl{1}     &   \qw      \\
				 \lstick{|\psi)}   &   \gate{Z}   &    \qw       \\			 	
			 }
	\caption{Circuit for $\hat{\Lambda}_{1}^{ql}\left(\hat{\sigma}_{z}\right)$.}
	\end{subfigure}
	\vspace{20pt}
	
	\begin{subfigure}[b]{0.45\columnwidth}
	~~~~~\Qcircuit @C=1.2em @R=1.3em {
				\lstick{|\psi)}   &   \ctrl{1}     &   \qw      \\
				 \lstick{\ket{\psi}}   &   \gate{Z}   &    \qw       \\			 	
			 }
	\caption{Circuit for $\hat{\Lambda}_{1}^{lq}\left(\hat{\sigma}_{z}\right)$.}
	\end{subfigure}
	\hfill
	\begin{subfigure}[b]{0.45\columnwidth}
	~~~~~\Qcircuit @C=1.2em @R=1.3em {
				\lstick{|\psi)}   &   \ctrl{1}     &   \qw      \\
				 \lstick{|\psi)}   &   \gate{Z}   &    \qw       \\			 	
			 }
	\caption{Circuit for $\hat{\Lambda}_{1}^{ll}\left(\hat{\sigma}_{z}\right)$.}
	\end{subfigure}
	\caption{Four different controlled-$Z$ gates}
	\label{fig:csigzall}
	\end{figure}

The operators in the last set,
$\hat{\Lambda}_{1}^{qq}\left(\hat{\sigma}_{z}\right)$, $\hat{\Lambda}_{1}^{ql}\left(\hat{\sigma}_{z}\right)$, $\hat{\Lambda}_{1}^{lq}\left(\hat{\sigma}_{z}\right)$ and $\hat{\Lambda}_{1}^{ll}\left(\hat{\sigma}_{z}\right)$, are four types of controlled-$Z$ operators with the 
same matrix representation
\begin{eqnarray}
	\left(\begin{array}{cccc}
		1 & 0 & 0 & 0 \\
		0 & 1 & 0 & 0 \\
		0 & 0 & 1 & 0 \\
		0 & 0 & 0 & -1 
		\end{array}\right)\,.
\end{eqnarray}
They differ from each other by the control and target bits being a qubit or a hybit as indicated by the superscript. 
Their circuits are shown in Fig.~\ref{fig:csigzall}.
The subscript indicates that there is only one  control bit in the gate. We will discuss gates with more than one
control bits in the Appendix. 

We have chosen the controlled-$Z$ gate instead of the control-NOT (CNOT) gate, which is more frequently used in quantum computing. The reason is that the CNOT gate may fail to be a complex Lorentz transformation. 
For example, we have 
\begin{eqnarray}
	\hat{\Lambda}^{ql}_{1}\left(\hat{\sigma}_{x}\right)^\dagger(\eta_{2,0}\otimes\eta_{1,1})\hat{\Lambda}^{ql}_{1}\left(\hat{\sigma}_{x}\right)
	\neq \eta_{2,0}\otimes\eta_{1,1}\,.
\end{eqnarray}
This indicates that the CNOT gate is not a complex Lorentz transformation when the control and target bits are qubit and hybit, respectively. However, it is ensured that the control-$Z$ gates are always complex Lorentz transformations whatever the control and target bits are.

\section{Application: Search Algorithm}
For the Lorentz computer  described in the previous section,
 we propose a search algorithm which is  faster than Grover's search algorithm for quantum computer.

We make use of  $n$ qubits for search along with one oracle qubit and one hybit. 
The task is to find out the target state $\ket{x}$ out of $2^n$ vectors $\ket{00\dots0} , \ket{00\dots1} , \dots , \ket{11\dots1}$,
which are stored in the $n$ qubits. The key in our algorithm is operator $\hat{Q} = \hat{O} \hat{\Lambda}_1^{ql}(\hat{V}) \hat{O}$ 
as shown in Fig. \ref{alg_cir}. The operator $\hat{O}$ is for oracle 
\be
\hat{O} = (\hat{I}-\ket{x}\bra{x}+\ket{x}\bra{x}(\ket{0_o}\bra{1_o}+\ket{1_o}\bra{0_o}))\,.
\ee
The control gate $\hat{\Lambda}_1^{ql}(\hat{V})$ 
has the oracle qubit as its control bit and the hybit as its target bit with the Lorentz transformation 
\be
\hat{V} = \left(\begin{array}{c c}
	\cosh\chi & \sinh\chi \\
	\sinh\chi & \cosh\chi 
\end{array}
\right)\,,
\ee
where $\chi$ is a positive constant . 
The algorithm goes as follows. 
\begin{enumerate}
\item Initialize the computer with Hadamard gates on each qubit (except the oracle qubit) 
\be
\ket{\Phi_0}\otimes \ket{0_o}\otimes |0) = \frac{1}{\sqrt{N}}\Big(\sum_{j = 0}^{2^n-1}\ket{j} \Big)\otimes \ket{0_o}\otimes |0)\,,
\ee
where $N=2^n$. 
\item Apply operator $\hat{Q}$ on the state vector for $k$ times, and get $\hat{Q}^k |\Phi_0)$

\item Measure the $n$ qubits and the hybit. 
\end{enumerate}
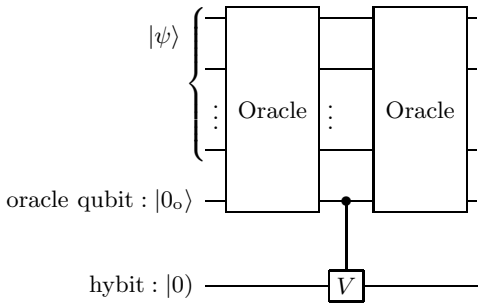
\begin{figure}[h!] 
	\centerline{
  \Qcircuit @C=0.4em @R=1.2em {
	 \lstick{} &   \qw  &   \multigate{4}{\rm Oracle}  &   \qw   &   \multigate{4}{\rm Oracle}     &   \qw     \\
	 \lstick{}  &   \qw &   \ghost{\rm Oracle}       &   \qw     &   \ghost{\rm Oracle}     &   \qw        \\
	 \lstick{}         &      \vdots   &    &       \lstick{\vdots }          &          &        \\
	 \lstick{}  &   \qw &   \ghost{\rm Oracle}       &   \qw           &   \ghost{\rm Oracle}    &   \qw      \inputgroupv{1}{4}{.8em}{.8em}{\ket{\psi}}\\
	 \lstick{\rm oracle\ qubit : \ket{0_o}}  &   \qw &   \ghost{\rm Oracle}       &   \ctrl{2}           &   \ghost{\rm Oracle}  &   \qw   \\
	  & & & & & \\
	 \lstick{{\rm hybit} : |0)}  &   \qw   &   \qw       &   \gate{V}           &   \qw     &   \qw         
  }
  }
	\caption{Circuit for operator $\hat{Q}$.}
	\label{alg_cir}
\end{figure}

The time complexity of our algorithm is $O(\log N)$ as we shall  prove below.
 Making use of $\hat{O}$ and $\hat{V}$, we get
\begin{eqnarray}
	&&\hat{Q}^k\left(\ket{\Phi_0}\otimes \ket{0_o}\otimes |0)\right)\nonumber\\
	&=& (\hat{I}-\ket{x}\bra{x}+\cosh(k\chi)\ket{x}\bra{x})\ket{\Phi_0}\otimes \ket{0_o}\otimes|0)\nonumber\\
	 &&+ (\sinh(k\chi)\ket{x}\bra{x})\ket{\Phi_0}\otimes \ket{0_o}\otimes|1)\,.
\end{eqnarray}
According to the theory of Lorentz mechanics, the probability of getting $\ket{x}$ is 
\begin{eqnarray}
P = \frac{\frac{1}{N}\cosh^2k\chi }{1-\frac{1}{N}+\frac{1}{N}\cosh^2k\chi}\,.
\end{eqnarray}
It is clear that $P\approx 1$ when $k\approx\frac{1}{\chi}\ln N$. So, 
the time complexity of our algorithm is $O(\log N)$.

For quantum computer, the Grover's algorithm can be simulated with Hamiltonians\cite{Farhi1998,Dam1,Cerf2000,Wilczek}. 
Similarly, our search algorithm can also be implemented with a Hamiltonian
\begin{eqnarray}
\hat{\mathcal H} = \ket{x}\bra{x}\otimes (\eta_{1,1}\hat{\mathcal H}_0)\,,
\end{eqnarray}
in which $e^{i\eta_{1,1}\hat{\mathcal H}_0} = \hat{V}$.

This Lorentz search algorithm with exponential speedup can be used to solve NP problems efficiently.
Any computation problem can be converted into a search problem of finding the proper solution out of all possible solutions.
For an NP problem, which  has $N=2^n$ possible solutions~\cite{Karp1972}, our algorithm requires to 
invoke the oracle $O(\log N) = O(n)$ times, while the time complexity for running the oracle is polynomial $O(n^p)$ by definition. 
As a result, the solution can be searched and found in a polynomial time scale with our Lorentz  
search algorithm. In other words, all NP problems  can be solved efficiently on a Lorentz computer. 
It is not clear whether NP-hard problems, such as maximum independent set problem~\cite{Xiao,Yu2021}, can be solved 
efficiently on a Lorentz computer. 

\section{Discussion and Conclusion} 
Our model of Lorentz computer may remind people of  standard quantum computer 
with postselection~\cite{Aaronson}.  In the final step, that only the components with $|0)$ are observable is 
equivalent to postselecting $|0)$ and discarding $|1)$. For convenience, we call it hyper-postselection. 
(1) Hyper-postselection is inspired by physics related to Bogoliubov-de Genne equation~\cite{wunjp,BEC,Artem}, 
not an arbitrary assumption; (2) It works only on hybits and does not work on qubits; (3) Due to the search algorithm presented
in the last section, it is clear that our Lorentz computer with hyper-postselection is at least as powerful as standard quantum computer 
with postselection. The search algorithm is the first algorithm that we found for the 
Lorentz computer; more powerful algorithms  may come up in the future.

There are other theoretical models of computer, which can also outperform the standard quantum computer. 
One is called digital memcomputing machine that is classical~\cite{memcomputing2015,zhang2021directed} 
and the other is called quantum computer with CTC (closed timelike curve)~\cite{bacon2004quantum}. 
They both allow dynamical evolution along a closed timelike world line, which has not been found to exist in nature.    
Adding nonlinearity to quantum mechanics may also speed up computing~\cite{abrams1998nonlinear}. 
However, this claim is very doubtful as nonlinearity in classical mechanics has never been found to speed up computing. 
The reason is that any nonlinear dynamics in a short time step can be approximated by linear gates. Nonlinear quantum 
dynamics can certainly be approximated with linear gates as well. In fact,  the effectiveness of quantum computing with CTC or nonlinearity has 
been seriously questioned~\cite{Bennett2009}.

In constructing our Lorentz computer, we have assumed that the Lorentz mechanics is fundamentally different from 
quantum mechanics in the spirit of how  Pauli discussed this kind of mechanics in 1940s~\cite{Pauli1943}.
However, as the Bogoliubov-de Gennes equation is an approximation of a more fundamental quantum equation~\cite{wunjp,BEC}, 
it is possible that our computer model can be constructed approximately in experiments. 
In fact, a Lorentz system was already demonstrated in a photon experiment with postselection~\cite{Li2019}. 

In sum,  we have set up a computational model based on  Lorentz mechanics. This model consists of both qubits
and hybits along with a set of universal logical gates. We have designed a search algorithm for Lorentz computer, 
which is exponentially faster than the Grover algorithm.  This explicitly shows that Lorentz computer is more powerful 
than quantum computer.

\acknowledgments
 We thank Qi Zhang for helpful discussion.  We are supported by the 
 National Key R\&D Program of China (Grants No.~2017YFA0303302, No.~2018YFA0305602), 
 the National Natural Science Foundation of China (Grant No. 11921005), and Shanghai 
 Municipal Science and Technology Major Project (Grant No.2019SHZDZX01).

\appendix
\section{Proof of University}
This appendix gives a  proof of universality for the following three sets of gates 
\ba
&&\{ \hat{H}, \hat{T}\}\,,~~\{ \hat{\tau}, \hat{T}\}\,,\nonumber\\
&&\{\hat{\Lambda}_{1}^{qq}\left(\hat{\sigma}_{z}\right), \hat{\Lambda}_{1}^{ql}\left(\hat{\sigma}_{z}\right), \hat{\Lambda}_{1}^{lq}\left(\hat{\sigma}_{z}\right) ,\hat{\Lambda}_{1}^{ll}\left(\hat{\sigma}_{z}\right)\}
\}\,.\nonumber
\ea
The mathematical meaning of universality means that any operator can be approximated to an arbitrary precision with the universal gates. 

Before laying out the proof in details in five steps, we point out that $\{ \hat{H} , \hat{T}\}$ is single qubit universal as already shown 
in quantum computation~\cite{nielson2000quantum,boykin1999universal}.

\subsection{From $\{\hat{\tau}, \hat{T}\}$ to arbitrary single hybit operators}
$\{\hat{\tau} , \hat{T}\}$ is single qubit universal, which means that any operator $\hat{\mathcal U}$ for single hybit  can be approximated by matrix product of a sequence of operator $\hat{\tau}$ and operator $\hat{T}$ , for example, $\hat{\tau}\hat{\tau}\hat{T}\hat{\tau}\cdots$. Error of the 
approximation $\vert\vert {\hat{\mathcal U}} - \hat{\tau}\hat{\tau}\hat{T}\hat{\tau}\cdots \vert\vert$ 
can be reduced to arbitrary small.

A single hybit operator is an element in group $U(1,1)$ or $SU(1,1)$ represented by a matrix~\cite{zhang2018lorentz}
\begin{eqnarray}
\left(\begin{array}{cc}
	\zeta & \gamma^*  \\
	\gamma & \zeta^*  \\
	\end{array}\right),
\quad
 \zeta \in \mathbb{C} , \gamma \in \mathbb{C}
\end{eqnarray}
or as  an exponential map
$
e^{-i\theta(i\hat{\sigma}_x n_x +i\hat{\sigma}_y n_y +\hat{\sigma}_z n_z )} = e^{i\theta \boldsymbol{n} \cdot \boldsymbol{\hat{\sigma}}}
$ in which $n_x,n_y,n_z \in \mathbb{R} , \boldsymbol{n} = (in_x,in_y,n_z) , \boldsymbol{\hat{\sigma}} = (\hat{\sigma}_x,\hat{\sigma}_y,\hat{\sigma}_z)$.  Note that we make no distinction between $U(m,n)$ and $SU(m,n)$ 
for the reason that an arbitrary overall phase is trivial. 

The elements in $SU(1,1)$ fall into three categories: space rotation, lightlike rotation, and pseudo rotation.
\begin{eqnarray}
&&e^{-i\theta(i\hat{\sigma}_x n_x +i\hat{\sigma}_y n_y +\hat{\sigma}_z n_z )}\nonumber\\
=&&\left\{
 \begin{array}{l}
 \cos\theta \hat{I} - i \sin\theta(i\hat{\sigma}_xn_x+i\hat{\sigma}_yn_y+\hat{\sigma}_zn_z) \\
 \quad \quad \quad \quad \quad \quad  \quad \quad \quad {\rm if} -n_x^2-n_y^2+n_z^2 = 1 \\
 \\
 \hat{I} - i \theta(i\hat{\sigma}_xn_x+i\hat{\sigma}_yn_y+\hat{\sigma}_zn_z)  \\
 \quad \quad \quad  \quad \quad \quad  \quad \quad \quad {\rm if} -n_x^2-n_y^2+n_z^2 = 0 \\
 \\
 {\rm \cosh}\theta \hat{I} - i {\rm \sinh}\theta(i\hat{\sigma}_xn_x+i\hat{\sigma}_yn_y+\hat{\sigma}_zn_z)  \\
 \quad \quad \quad  \quad \quad \quad  \quad \quad \quad {\rm if}  -n_x^2-n_y^2+n_z^2 = -1  \\
 \end{array}
 \right.
\end{eqnarray}

It is easy to verify that the operators $\{\hat{\tau} , \hat{T}\}$  are all space rotation.
But surprisingly, not only can we generate all the space rotations in $U(1,1)$ with them,
 but also all the lightlike rotations and pseudo rotations.

Consider an operator 
\begin{eqnarray}
\hat{P} &&= \hat{T}\hat{\tau} \nonumber\\
&&= \sqrt{2}\sin\frac{\pi}{8}\hat{I}+i(i\hat{\sigma}_x\cos\frac{\pi}{8}+i\hat{\sigma}_y\sin\frac{\pi}{8}+\sqrt{2}\hat{\sigma}_z\cos\frac{\pi}{8}) \nonumber\\
&&= \cos\theta_0I+i\sin\theta_0(i\hat{\sigma}_xn_x+i\hat{\sigma}_yn_y+\hat{\sigma}_zn_z)\,.
\end{eqnarray}
in which $\theta_0= {\arccos(\sqrt{2}\sin\frac{\pi}{8})}$.
It can be proved that 
$\frac{\theta_0}{\pi}$
is an irrational number
according to the theory of cyclotomic polynomial~\cite{dummit2004abstract} by noticing the fact 
that the minimal polynomial of $e^{i\theta_0}$ in $\mathbb{Z}[n]$, $x^8-4x^6-2x^4-4x+1$, 
 is not a cyclotomic polynomial.
This means that, with an appropriate integer $n$,
$
\hat{P}^n = \cos n\theta_0I+i\sin n\theta_0(i\hat{\sigma}_xn_x+i\hat{\sigma}_yn_y+\hat{\sigma}_zn_z)
$
can approximate the space rotation along spacelike axis $\boldsymbol{n}_1=(in_x,in_y,n_z)$ for an arbitrary angle with arbitrary precision.
Thus we get space rotation 
$e^{i\theta \boldsymbol{n}_1 \cdot \boldsymbol{\hat{\sigma}}}$
for arbitrary $\theta$.

Applying similarity transformations to this operator, we can obtain another two space rotations
\begin{eqnarray}
    \hat{T} e^{i\theta_1 \boldsymbol{n}_1 \cdot \boldsymbol{\hat{\sigma}}} \hat{T}^{\dagger} =  e^{i\theta_1 \boldsymbol{n}_2 \cdot \boldsymbol{\hat{\sigma}}}
\end{eqnarray}
\begin{eqnarray}
    \hat{T}^2 e^{i\theta_2 \boldsymbol{n}_1 \cdot \boldsymbol{\hat{\sigma}}} (\hat{T}^2)^\dagger = e^{i\theta_2 \boldsymbol{n}_3 \cdot \boldsymbol{\hat{\sigma}}}
\end{eqnarray}
  for arbitrary $\theta$ along axis $\boldsymbol{n}_2$ and $\boldsymbol{n}_3$.
Noticing the linearly independence of $\boldsymbol{n}_1$, $\boldsymbol{n}_2$ and $\boldsymbol{n}_3$,
 we can write any vector $\theta \boldsymbol{n}$ into the linear superposition of them
 \begin{eqnarray}
	\theta \boldsymbol{n} = \alpha_1 \boldsymbol{n}_1 + \alpha_2 \boldsymbol{n}_2 + \alpha_3 \boldsymbol{n}_3.
 \end{eqnarray}
So, any single hybit operator can be written as
\be
e^{i\theta \boldsymbol{n} \cdot \boldsymbol{\hat{\sigma}}} = e^{i(\alpha_1 \boldsymbol{n}_1 \cdot \boldsymbol{\hat{\sigma}}+\alpha_2 \boldsymbol{n}_2 \cdot \boldsymbol{\hat{\sigma}}+\alpha_3 \boldsymbol{n}_3 \cdot \boldsymbol{\hat{\sigma}})}\,.
\ee
Applying Baker-Hausdorff formula with a large enough integer $\ell$
\begin{eqnarray}
e^{i\theta \boldsymbol{n} \cdot \boldsymbol{\hat{\sigma}}} \approx (e^{i\frac{\alpha_1}{\ell} \boldsymbol{n}_1 \cdot \boldsymbol{\hat{\sigma}}}e^{i\frac{\alpha_2}{\ell} \boldsymbol{n}_2 \cdot \boldsymbol{\hat{\sigma}}}e^{i\frac{\alpha_3}{\ell} \boldsymbol{n}_3 \cdot \boldsymbol{\hat{\sigma}}})^\ell
\end{eqnarray}
we can see that $e^{i\theta \boldsymbol{n} \cdot \boldsymbol{\hat{\sigma}}}$ is written in the form of product of rotations along $\boldsymbol{n}_1 , \boldsymbol{n}_2 $, or $\boldsymbol{n}_3$, which we already know how to generate.

\subsection{From controlled-$Z$ gate $\hat{\Lambda}_{1}\left(\hat{\sigma}_z\right)$ to controlled-$V$ gate $\hat{\Lambda}_{1}(\hat{V})$}
\label{sec:greetings}
For simplicity, we shall omit the superscripts and write
 $\hat{\Lambda}^{qq}_k(\hat{v})$, $\hat{\Lambda}^{ql}_k(\hat{v})$, $\hat{\Lambda}^{lq}_k(\hat{v})$, or 
 $\hat{\Lambda}^{ll}_k(\hat{v})$ as $\hat{\Lambda}_k(\hat{v})$ 
when the control and target bits are clear, or no confusion would arise. Here $k$ denotes the number of 
the control bits. 

We are going to construct the controlled-$V$ gate, $\hat{\Lambda}_{1}(\hat{V})$,
 where $\hat{V}$ denotes an arbitrary unitary operator when the target bit is a qubit, or
 an arbitrary complex Lorentz operator when the target bit is a hybit.
 If the target bit is qubit, corresponding to Fig.~\ref{fig:csigzall}(a,c),
 the proof is exactly the same as in quantum computing~\cite{nielson2000quantum,boykin1999universal}.
We only need to give the proof when the target bit is a hybit, as shown in Fig.~\ref{fig:csigzall}(b,d).
For these two cases,  the metric is $\operatorname{diag}(1,-1,1,-1)$ or $\operatorname{diag}(1,-1,-1,1)$.

With the following similarity transformation 
\be
(\hat{I} \otimes \hat{V}) \hat{\Lambda}_{1}\left(\hat{\sigma}_z\right) (\hat{I} \otimes \hat{V})^{-1}
=\hat{\Lambda}_{1}(\hat{V} \hat{\sigma}_z \hat{V}^{-1})  \,,
\ee
it is clear that when $\hat{V}$ goes through all the single hybit operators,
 $\hat{\Lambda}_{1}\left(\boldsymbol\hat{\sigma} \cdot \boldsymbol{n}\right)$ 
 for arbitrary $\boldsymbol{n}\cdot \boldsymbol{n} = 1$ can be generated.
We consider three specific such operators 
$\hat{\Lambda} ( \hat{\sigma}_z )$, $\hat{\Lambda} ( \hat{\sigma}_z \cosh\alpha + i \hat{\sigma}_y \sinh \alpha )$, 
$\hat{\Lambda} ( \hat{\sigma}_z \cosh\beta + i\hat{\sigma}_x \sinh\beta)$, 
in which $\alpha$ and $\beta$ are two unequal and nonzero real number.
The product of these three operators is
\begin{eqnarray}
\hat{P'}
=&&\hat{\Lambda}_1(\hat{\sigma}_z) \hat{\Lambda}_1(\hat{\sigma}_z \cosh\alpha + i \hat{\sigma}_y \sinh \alpha) \nonumber\\
&& \hat{\Lambda}_1(\hat{\sigma}_z \cosh\beta + i\hat{\sigma}_x \sinh\beta)\nonumber\\
= &&
 \hat{\Lambda}_1(
 i(\sinh\alpha \sinh\beta \hat{I} 
 -i (\cosh\alpha \cosh \beta \hat{\sigma}_z \nonumber\\
 &&+ i \cosh \alpha \sinh \beta \hat{\sigma}_x -i \sinh\alpha \cosh\beta \hat{\sigma}_y))
 )
 \nonumber\\
= &&
 \hat{\Lambda}_1(
 i e^{-i \arccos{(\sinh\alpha \sinh\beta)}\boldsymbol{\hat{\sigma}}\cdot\boldsymbol{n}}
 )
\end{eqnarray}
with
\be
\boldsymbol{n} = (i \cosh\alpha \sinh\beta , -i \sinh\alpha \cosh\beta , \sinh\alpha \sinh\beta)\,.
\ee
If we choose
 $\alpha , \beta$
 such that $\sinh\alpha \cosh\beta < 1$
 and $\frac{\arccos{(\sinh\alpha \sinh\beta)}}{\pi}$ is irrational,
 we can generate $\hat{\Lambda}_1(e^{i\theta \boldsymbol{\hat{\sigma}}\cdot\boldsymbol{n}})$ for arbitrary $\theta$ 
 with $\hat{P'}^k$. Then we can generate $\hat{\Lambda}_1(\hat{V})$ for all $\hat{V} \in SU(1,1)$ using the same trick as the previous section.

And we notice that 
\begin{eqnarray}
\left(\begin{array}{cc}
	1 &0 \\
	0 & e^{i\phi}
\end{array}\right)
\otimes \hat{I}
=
\left(\begin{array}{cccc}
	1 &0 &0 &0\\
	 0 & 1 &0 & 0\\
	 0& 0& e^{i\phi} &0 \\
	 0& 0& 0& e^{i\phi} \\
\end{array}\right)
= \hat{\Lambda}_1(e^{i\phi})\quad\quad\,.
\end{eqnarray}
So we have $\hat{\Lambda}_1(e^{i\phi}\hat{V}) = \hat{\Lambda}_1(e^{i\phi}) \hat{\Lambda}_1(\hat{V})$, and thus obtain $\hat{\Lambda}_1(\hat{V}_0)$ for arbitrary $\hat{V}_0 \in U(1,1)$

\subsection{From controlled-$V$ gate $\hat{\Lambda}_{1}\left(\hat{V}\right)$ to controlled-$V$ gate 
$\hat{\Lambda}_{k}(\hat{V})$ with $k$ control bits}
We denote controlled-$V$ gate with $k$ control bits as $\hat{\Lambda}_{k}(\hat{V})$,
 and we are going to generate $\hat{\Lambda}_{k}\left(\hat{\sigma}_{z}\right)$ for arbitrary positive integer $k$
 and arbitrary type of control and target bits(qubit or hybit). The following proof holds whether the bits are qubits, hybits or both, so the superscript of $\hat{\Lambda}_{k}(\hat{V})$ is omitted.\\

\begin{figure}[h!] 
	\centerline{
  		\Qcircuit @C=0.8em @R=0.75em {
			   & \qw & \qw  &   \qw  & \ctrl{2} & \qw & \qw & & & & \qw & \multigate{2}{\hat{W}_3(\hat{V} \hat{\sigma}_z \hat{V}^{-1})} & \qw\\	 
			   & \qw & \qw  &   \ctrl{1}  & \qw & \qw & \qw & & = & & \qw & \ghost{\hat{W}_3(\hat{V} \hat{\sigma}_z \hat{V}^{-1})} & \qw\\
			   & \qw & \gate{{V}^{-1}} &   \gate{Z}  & \gate{Z} & \gate{{V}} & \qw & & & & \qw & \ghost{\hat{W}_3(\hat{V} \hat{\sigma}_z \hat{V}^{-1})} & \qw\\
	 	}
  	}
	\caption{Circuit for operator $\hat{W}_3(\hat{V} \hat{\sigma}_z \hat{V}^{-1})$.}
	\label{fig:opU}
\end{figure}
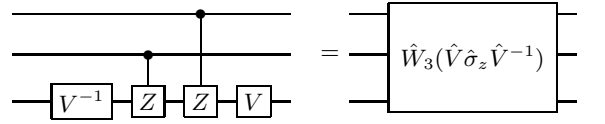

\begin{figure}[h!] 
	\centerline{
  		\Qcircuit @C=0.8em @R=0.75em {
			    & \qw  &   \qw  & \ctrl{2} & \multigate{2}{\hat{W}_3(\hat{\sigma}_z^{-\frac{1}{2}})} & \qw & & & & \qw & \ctrl{2} & \qw\\	 
			    & \qw  &   \ctrl{1}  & \qw & \ghost{\hat{W}_3(\hat{\sigma}_z^{-\frac{1}{2}})} & \qw & & = & & \qw & \ctrl{1} & \qw\\
			    & \qw &   \gate{Z^{\frac{1}{2}}}  & \gate{Z^{\frac{1}{2}}} & \ghost{\hat{W}_3(\hat{\sigma}_z^{-\frac{1}{2}})} & \qw & & & & \qw & \gate{Z} & \qw\\
	 	}
	  }
	\caption{Circuit for operator $\hat{\Lambda}_2(\hat{\sigma}_z)$.}
	\label{fig:ccsigz}
\end{figure}
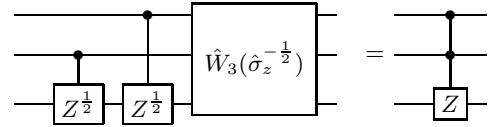

 We are first to construct $\hat{\Lambda}_2(\hat{V})$ for arbitrary $\hat{V}$. 
 $\hat{\Lambda}_2(\hat{V})$ is an $8\times 8$ matrix, which we denote as a $4\times 4$ matrix 
 with each element being a $2\times 2$  matrix.  Consider operator $\hat{W}_3(\hat{V} \hat{\sigma}_z \hat{V}^{-1})$,
 whose  circuit representation is shown in Fig.~\ref{fig:opU}. The subscript  indicates the number of relevant qubits and hybits is 3.
\be
	\hat{W}_3(\hat{V} \hat{\sigma}_z \hat{V}^{-1})
 = 
\left(\begin{array}{cccc}
	\hat{I}& 0&0 &0 \\
	 0&\hat{V} \hat{\sigma}_z \hat{V}^{-1} & 0&0 \\
	 0&0 &\hat{V} \hat{\sigma}_z \hat{V}^{-1}&0 \\
	 0&0 & 0&\hat{I} \\
\end{array}
\right)
\ee
\\
\\
The form of operator $\hat{W}_3(\hat{V} \hat{\sigma}_z \hat{V}^{-1})$ is similar to operator $\hat{\Lambda}_1(\hat{V} \hat{\sigma}_z \hat{V}^{-1})$ in the previous section,
 then we can generate 
 \begin{eqnarray}
\hat{W}_3(\hat{V}_0) = 
 \left(\begin{array}{cccc}
	\hat{I}&0 &0 &0 \\
	 0&\hat{V_0} &0 &0 \\
	 0& 0&\hat{V_0} &0 \\
	 0&0 &0 &\hat{I} \\
\end{array}
\right)
\end{eqnarray}
for arbitrary $V_0$ using the same technique. Specially we can generate
\begin{eqnarray}
\hat{W}_3(\hat{\sigma}_z^{-\frac{1}{2}}) = \left(\begin{array}{cccc}
	\hat{I}&0 &0 &0 \\
	0 & \hat{\sigma}_z^{-\frac{1}{2}} & 0&0 \\
	0 &0 & \hat{\sigma}_z^{-\frac{1}{2}} &0 \\
	 0 & 0 &0 &\hat{I} \\
\end{array}
\right)\,
\end{eqnarray}
with which we can construct $\hat{\Lambda}_2(\hat{\sigma}_z)$ with the circuit in Fig.~\ref{fig:ccsigz}.

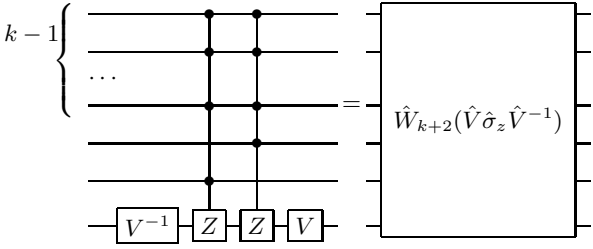
\begin{figure}[h!] 
	\centerline{
  		\Qcircuit @C=0.6em @R=0.65em {  
			& & \qw & \qw  &   \ctrl{6}  & \ctrl{6} & \qw & \qw & &  & \multigate{6}{\hat{W}_{k+2}(\hat{V} \hat{\sigma}_z \hat{V}^{-1})} & \qw\\
			& & \qw & \qw  &   \ctrl{5}  & \ctrl{5} & \qw & \qw & &  & \ghost{\hat{W}_{k+2}(\hat{V} \hat{\sigma}_z \hat{V}^{-1})} & \qw\\
			&	& \cdots &   &     & &  &  & &  & &  &  & \\
			&	& \qw & \qw  &   \ctrl{3}  & \ctrl{3} & \qw & \qw & = &  &   \ghost{\hat{W}_{k+2}(\hat{V} \hat{\sigma}_z \hat{V}^{-1})} & \qw
				\inputgroupv{1}{4}{.8em}{.8em}{ k-1 }\\	 
			&   & \qw & \qw  &   \qw  & \ctrl{2} & \qw & \qw & & &   \ghost{\hat{W}_{k+2}(\hat{V} \hat{\sigma}_z \hat{V}^{-1})} & \qw\\	 
			&   & \qw & \qw  &   \ctrl{1}  & \qw & \qw & \qw & &  &  \ghost{\hat{W}_{k+2}(\hat{V} \hat{\sigma}_z \hat{V}^{-1})} & \qw\\
			&   & \qw & \gate{{V}^{-1}} &   \gate{Z}  & \gate{Z} & \gate{{V}} & \qw & & &  \ghost{\hat{W}_{k+2}(\hat{V} \hat{\sigma}_z \hat{V}^{-1})} & \qw\\
	 	}
  	}
	\caption{$k$-control-bit version of Fig.~\ref{fig:opU}.}
	\label{fig:conn1}
\end{figure}

\begin{figure}[h!] 
	\centerline{
  		\Qcircuit @C=0.8em @R=0.75em {
				  &   \ctrl{3}  & \ctrl{6} & \multigate{6}{\hat{W}_{k+2}(\hat{\sigma}_z^{-\frac{1}{2}})} & \qw & & & & \qw & \ctrl{2} & \qw\\	
				 &   \ctrl{3}  & \ctrl{3} & \ghost{\hat{W}_{k+2}(\hat{\sigma}_z^{-\frac{1}{2}})} & \qw & & & & \qw & \ctrl{2} & \qw\\ 
				&  \cdots &     &  &  &  & & & & \cdots &  & \\	 
				  &   \ctrl{3}  & \ctrl{3} & \ghost{\hat{W}_{k+2}(\hat{\sigma}_z^{-\frac{1}{2}})} & \qw & & & & \qw & \ctrl{2} & \qw
				\inputgroupv{1}{4}{.8em}{.8em}{k-1  \quad}\\	 
			      &   \qw  & \ctrl{2} & \ghost{\hat{W}_{k+2}(\hat{\sigma}_z^{-\frac{1}{2}})} & \qw & & = & & \qw & \ctrl{2} & \qw\\	 
			     &   \ctrl{1}  & \qw & \ghost{\hat{W}_{k+2}(\hat{\sigma}_z^{-\frac{1}{2}})} & \qw & & & & \qw & \ctrl{1} & \qw\\
			     &   \gate{Z^{\frac{1}{2}}}  & \gate{Z^{\frac{1}{2}}} & \ghost{\hat{W}_{k+2}(\hat{\sigma}_z^{-\frac{1}{2}})} & \qw & & & & \qw & \gate{Z} & \qw\\
	 	}
	  }
	\caption{$k$-control-bit version of Fig.~\ref{fig:ccsigz}.}
	\label{fig:conn2}
\end{figure}
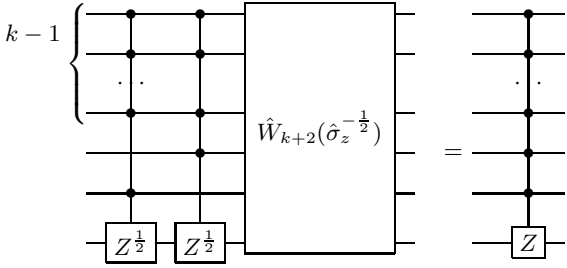

We have already obtained operator $\hat{\Lambda}_2(\hat{\sigma}_z)$ using $\hat{\Lambda}_1(\hat{\sigma}_z)$ and 
single qubit or single hybit operators. Applying the same steps, we can generate $\hat{\Lambda}_{k+1}(\hat{\sigma}_z)$ 
using $\hat{\Lambda}_{k}(\hat{\sigma}_z)$ and single qubit or hybit operators, as 
described in Fig.~\ref{fig:conn1} and Fig.~\ref{fig:conn2}. 
With $\hat{\Lambda}_{k}(\hat{\sigma}_z)$, $\hat{\Lambda}_{k+1}(\hat{V})$ can be straightforwardly 
generated in the same way as generating $\hat{\Lambda}_{1}(\hat{V})$ with $\hat{\Lambda}_{1}(\hat{\sigma}_z)$.
\\

\subsection{Arbitrary operators can be factorized into two-level matrices}
Similar to standard quantum computing[\onlinecite{nielson2000quantum}],
 two-level matrix can be defined in $U(m,n)$. We denote it as $\hat{b}_{i,j}(\hat{V})$, 
 in which $ 1\leq i < j \leq m+n$ and 
 \begin{eqnarray}
	 \hat{V} = \left(
		\begin{array}{cc}
			V_{11}& V_{12}\\
			V_{21} &V_{22}
		\end{array}
	\right)
 \end{eqnarray}
is a unitary or complex Lorentz matrix.
$\hat{b}_{i,j}(\hat{V})$ can be represented in the form of matrix
 \begin{eqnarray}
	\begin{array}{c}
		\\
		\\
		\\
		{i{\rm -th}~{\rm row}}\\
		\\
		\\
		\\
		{j{\rm -th}~{\rm row}}\\
		\\
		\\
		\\
	\end{array}
	\left(\begin{array}{ccccccccccc}
		1& \cdots&0 &0 &0 &\cdots &0 &0 &0 &\cdots &0\\
		\vdots& \ddots&\vdots &\vdots &\vdots &\vdots&\vdots &\vdots&\vdots &\vdots &\vdots\\
		0& \cdots&1 &0 &0 &\cdots &0 &0 &0 &\cdots &0\\
		0&\cdots & 0& V_{11} &0 &\cdots &0  & V_{12} &0 &\cdots &0\\
		0&\cdots & 0& 0& 1& \cdots &0  &0 &0 &\cdots &0\\
		\vdots&\vdots &\vdots & \vdots&\vdots & \ddots & \vdots &\vdots &\vdots &\vdots &\vdots \\
		0&\cdots  & 0& 0&0 &\cdots  & 1  &0 & 0& \cdots&0\\
		0&\cdots &0 &  V_{21}&0 &\cdots  &0   & V_{22}& 0& \cdots&0\\
		0&\cdots  & 0& 0&0 &\cdots  & 0 &0 & 1& \cdots&0\\
		\vdots&\vdots &\vdots & \vdots& \vdots& \vdots &\vdots  &\vdots &\vdots & \ddots &\vdots\\
		0& \cdots&0 &0 &0 &\cdots &0 &0 &0 &\cdots &1 \\
	\end{array}
	\right)\nonumber\\
 \end{eqnarray}
When restricted on the subspace spanned by $\{ \ket{i} , \ket{j} \}$, 
we simply have $\hat{b}_{i,j}(\hat{V})=\hat{V}$.  As $\hat{b}_{i,j}(\hat{V}) \in U(m,n)$, 
$\hat{V}$ is a complex Lorentz matrix matrix when $1\leq i \leq m < j \leq m+n$, or a unitary matrix otherwise.
When restricted on the orthogonal complement space $\hat{b}_{i,j}(\hat{V})$ is an identity matrix.

For any operator $A\in U(m,n)$ in which $m \ge 2$ and $n \ge 2$, consider $A^{(1)} = \hat{b}_{1,2}(\hat{V})A$ ,
which is in the form of  matrix 
 \begin{eqnarray}
 \left(
 \begin{array}{c}
	 A^{(1)}_{1,1}\\
	 A^{(1)}_{2,1}
 \end{array}
 \right)
 =
 \left(
 \begin{array}{cc}
	V_{11}& V_{12}\\
	- V_{12}^* & V_{11}^*
\end{array}
\right)
\left(
\begin{array}{c}
	A_{1,1}\\
	A_{2,1}
\end{array}
\right)
\end{eqnarray}
in which $A_{k,l}$ and $A^{(1)}_{k,l}$ denote the $k$-th row $l$-th column element of matrix $A$ and $A^{(1)}$, 
respectively ($A^{(2)}_{k,l}$ $A^{(3)}_{k,l}$ in the following text are similar).
We can choose appropriate $\hat{V}$ such that $A_{2,1}^{(1)} = 0$.
We continue this procedure with  $\hat{b}_{1,j}$ ($j=3,4,\cdots,m$) and obtain  $A^{(2)}= \hat{b}_{1,m}\cdots\hat{b}_{1,3}A^{(1)}$  
such that $A^{(2)}_{2,1} = \cdots = A^{(2)}_{m,1} = 0$.  
To make $A_{j,1}=0$ with $j>m+1$,  we can use a different set of $\hat{b}_{i,j}$ and 
obtain $A^{(3)}=\hat{b}_{m+1,m+n}\cdots\hat{b}_{m+1,m+2}A^{(2)}$ such 
 that $A^{(3)}_{2,1} = \cdots = A^{(3)}_{m,1} = A^{(3)}_{m+2,1} = \cdots = A^{(3)}_{m+n,1} = 0$. 
 To make $A_{m+1,1}=0$, we use $A^{(4)}=\hat{b}_{1,m+1}(\hat{V}')A^{(3)}$,
 \begin{eqnarray}
 \left(
 \begin{array}{c}
	 A^{(4)}_{1,1}\\
	 A^{(4)}_{m+1,1}
 \end{array}
 \right)
 =
 \left(
 \begin{array}{cc}
	V'_{11}& V'_{12}\\
	{V'}_{12}^* & {V'}_{11}^*
\end{array}
\right)
\left(
\begin{array}{c}
	A^{(3)}_{1,1}\\
	A^{(3)}_{m+1,1}
\end{array}
\right)
\end{eqnarray}
By choosing appropriate values for the elements of $\hat{V}'$,  
we can make $A^{(4)}_{1,1} = 1$ and $A^{(4)}_{m+1,1} = 0$, for the reason that
 \begin{eqnarray}
 \absolutevalue{A^{(3)}_{1,1}}^2 - \absolutevalue{A^{(3)}_{m+1,1}}^2
 = \sum_{i=1}^{m}\absolutevalue{A^{(3)}_{i,1}}^2 - \sum_{i=m+1}^{m+n}\absolutevalue{A^{(3)}_{i,1}}^2 
 = 1\nonumber\\
\end{eqnarray}
We now have $A_{1,2}=\cdots=A_{1,m+n}=0$. As column vectors in matrix $A^{(4)}$ are orthogonal to each other, 
  we have $A_{2,1}=\cdots=A_{m+n,1}=0$. 
  Thus we have reduced matrix $A$ into a block matrix 
\begin{eqnarray}
\left(\begin{array}{c|ccc}
1 & 0 & \cdots & 0 \\
\hline 0 & & \\
\vdots & & U(m-1,n) &\\
0 &&
\end{array}\right)
\end{eqnarray}
By induction, we can reduce $A$ into an identity matrix $\hat{I}$ using a series of two-level matrices.  
This is equivalent to the assertion that arbitrary operators $A$  can be factorized into a series of two-level matrices.

\subsection{From  $\hat{\Lambda}_{N-1}(\hat{V})$ to two-level matrices}
In this section, our goal is to construct all the two-level matrices using $\hat{\Lambda}_{N-1}(\hat{V})$, where
$N$ is the total number of qubits and hybits. The following proof holds whether the bits are qubits, hybits, or both, 
so the superscripts of $\hat{\Lambda}_{k}(\hat{V})$ are omitted.

We notice that $\hat{\Lambda}_{N-1}(\hat{V})$ can be considered as a special kind of two-level matrices $\hat{b}_{i,j}$ 
where only one  qubit (or hybit) between $\ket{i}$ and $\ket{j}$ is different, while the 
rest of qubits (or hybits) are all $\ket{1}$ (or $|1)$), for example, 
\ba
\ket{i}& =& \ket{1,1,0,1,1}\otimes|1,1),\\
\ket{j} &= &\ket{1,1,1,1,1}\otimes|1,1)\,.
\ea

Different from $\hat{\Lambda}_{N-1}(\hat{V})$, $\hat{b}_{i,j}$ of which the Hamming distance between $i$ and $j$ is 1 also requires only one different qubit(or hybit) between $\ket{i}$ and $\ket{j}$ but imposes no constraint on the rest of the qubits and hybits, for example,
\ba
\ket{i}& =& \ket{1,1,0,0,0}\otimes|1,0),\\
\ket{j} &= &\ket{1,1,1,0,0}\otimes|1,0)\,.
\ea
For the case of $N=2$, 
only $\hat{b}_{i,j}(\hat{\sigma}_z)$ with $\ket{i}=\ket{00}$, $\ket{j} = \ket{01}$ is not a controlled-$Z$ gate $\Lambda_1(\sigma_z)$. 
It can be constructed as 
\be
    \hat{b}_{i,j}(\hat{\sigma}_z) = \hat{\Lambda}_1(\hat{\sigma}_z)(\hat{I}\otimes \hat{\sigma}_z)
	\label{eq:bij}
\ee
Noticing that $\hat{b}_{i,j}$ with the Hamming distance between $\ket{i}$ and $\ket{j}$ equal to 1 can be obtained by exchanging $\ket{0}$ and $\ket{1}$(or $|0)$ and $|1)$) for some qubits (hybits),
we can obtain such $\hat{b}_{i,j}(\hat{V})$ for $N > 2$ by applying steps in Appendix~\ref{sec:greetings} 
but substitute some $\hat{\Lambda}_1(\hat{\sigma}_z)$ operators with the operator in Eq.(\ref{eq:bij}),
as the substitution essentially exchange $\ket{0}$ and $\ket{1}$(or $|0)$ and $|1)$) for the control bits.

Now we are going to generate $\hat{b}_{i,j}$ of which the Hamming distance between $i$ and $j$ is 2.
 We can always find a basis vector $\ket{k}$ such that the Hamming distance between $i,k$ and $j,k$ are both 1,
 and we are to generate $\hat{b}_{i,j}$ by $\hat{b}_{i,k}$ and $\hat{b}_{j,k}$.
Considering the subspace spanned by $\{\ket{i}, \ket{j}, \ket{k}\}$,
 by symmentry, we only need to consider three kinds of metric : $\operatorname{diag}(1,1,1)$ , $\operatorname{diag}(-1,1,1)$ and $\operatorname{diag}(1,1,-1)$.

For each case, we construct $\hat{b}_{i,j}$ with $\hat{b}_{i,k}$ and $\hat{b}_{j,k}$ in the form of matrix restricted on the subspace.
If the metric is $\operatorname{diag}(1,1,1)$
\be
\left(
\begin{array}{ccc}
	\zeta & \gamma & 0\\
	-\gamma^* & \zeta^* & 0\\
	0 & 0& 1 \\
\end{array}
\right)
=
\left(
\begin{array}{ccc}
	1 & 0 & 0\\
	0  & 0 & 1 \\
	 0& 1 & 0 \\
\end{array}\right)
\left(
\begin{array}{ccc}
	\zeta & 0 & \gamma \\
	 0 & 1 &  0\\
	 - \gamma^* & 0 & \zeta^* \\
\end{array}\right)
\left(
\begin{array}{ccc}
	1 & 0 & 0\\
	 0 & 0 & 1 \\
	0 & 1 & 0 \\
\end{array}\right)\,.
\ee
If the metric is $\operatorname{diag}(-1,1,1)$,
\be
\left(
\begin{array}{ccc}
	\zeta & \gamma & 0\\
	\gamma^* & \zeta^* & 0\\
	0 & 0& 1 \\
\end{array}
\right)
=
\left(
\begin{array}{ccc}
	1 & 0 & 0\\
	 0 & 0 & 1 \\
	0 & 1 & 0 \\
\end{array}\right)
\left(
\begin{array}{ccc}
	\zeta & 0 & \gamma \\
	0  & 1 &  0\\
	 \gamma^* & 0 & \zeta^* \\
\end{array}\right)
\left(
\begin{array}{ccc}
	1 & 0 & 0\\
	0  & 0 & 1 \\
	0 & 1 &  0\\
\end{array}\right)\,.
\ee
If the metric is $\operatorname{diag}(1,1,-1)$,
\begin{eqnarray}
&&\left(
\begin{array}{ccc}
	\zeta & \gamma &0 \\
	-\gamma^* & \zeta^* & 0\\
	0 &0 & 1 \\
\end{array}
\right)
= \nonumber\\
&&
\left(
\begin{array}{ccc}
	\frac{\sqrt{1+\absolutevalue{\gamma}^2}}{\zeta^*} & 0 & -\sqrt{2}\frac{\gamma}{\zeta^*} \\
	0  & 1 & 0 \\
	  -\sqrt{2}\frac{\gamma^*}{\zeta} & 0 & \frac{\sqrt{1+\absolutevalue{\gamma}^2}}{\zeta} \\
\end{array}\right)
\left(
\begin{array}{ccc}
	1 & 0 & 0\\
	0  & \sqrt{2} & -1 \\
	0 & -1 & \sqrt{2} \\
\end{array}\right)
\nonumber\\
&&
\left(
\begin{array}{ccc}
	\sqrt{1+\absolutevalue{\gamma}^2} & 0 & \gamma \\
	0  & 1 &  0\\
	 \gamma^* & 0 & \sqrt{1+\absolutevalue{\gamma}^2} \\
\end{array}\right)
\left(
\begin{array}{ccc}
	1 & 0 & 0\\
	0 & \frac{\sqrt{2}}{\zeta} & \frac{\sqrt{1+\absolutevalue{\gamma}^2}}{\zeta^*} \\
	0 & \frac{\sqrt{1+\absolutevalue{\gamma}^2}}{\zeta} & \frac{\sqrt{2}}{\zeta^*} \\
\end{array}
\right)\nonumber\\
\end{eqnarray}
In above identities, $\zeta$ and $\gamma$ are arbitrary complex numbers.
By induction, we can generate $\hat{b}_{i,j}$ of which the Hamming distance between $i$ and $j$ is arbitrary. 
This concludes the proof. 


%

\end{document}